\documentstyle[epsfig,times]{mn}
\def\spose#1{\hbox to 0pt{#1\hss}}
\def\oo{[O\textsc{ii}]\,\,}
\def\oohb{[O\textsc{ii}]/H$\beta$\,\,}

\def\lta{\mathrel{\spose{\lower 3pt\hbox{$\mathchar"218$}}\raise 2.0pt\hbox{$\mathchar"13C$}}}
\def\gta{\mathrel{\spose{\lower 3pt\hbox{$\mathchar"218$}}\raise 2.0pt\hbox{$\mathchar"13E$}}}
\def\arcsec{$^{\prime\prime}$}
\def\crat{C\textsc{iii}]1909/C\textsc{ii}]2326 }
\def\nrat{[Ne\textsc{iii}]3869/[Ne\textsc{v}]3426 }

\title[Spectroscopy of $\mathbf{z \sim 1}$ 6C radio galaxies - II]
{Deep spectroscopy of $\mathbf{z \sim 1}$ 6C radio galaxies - II. Breaking the
redshift--radio power degeneracy}
      
\author[K.\,J.\, Inskip {\it et al.}]
{K.\,J.\, Inskip$^1$\footnotemark, P.\,N.\,Best$^2$,
H.\,J.\,A.\,R\"{o}ttgering$^3$, S.\,Rawlings$^4$, G.\,Cotter$^1$,\cr 
and M.\,S.\,Longair$^1$\\ 
$^1$ Cavendish Laboratory, Madingley Road, Cambridge, CB3 0HE,\\ $^2$
Institute for Astronomy, Royal Observatory Edinburgh, Blackford Hill,
Edinburgh, EH9 3HJ\\$^3$ Sterrewacht Leiden, Postbus 9513, 2300 RA
Leiden, the Netherlands\\ $^4$ Department of Astrophysics, University
of Oxford, Keble Road, Oxford OX1 3RH\\ }

\date{}

\pagerange{\pageref{firstpage}--\pageref{lastpage}}

\pubyear{2002}

\begin{document}

\label{firstpage}

\maketitle

\begin{abstract}
The results of a spectroscopic analysis of 3CR and 6C radio galaxies
at redshift $z \sim 1$ are contrasted with the properties of lower
redshift radio galaxies, chosen to be matched in radio luminosity to
the 6C sources studied at $z \sim 1$, thus enabling the
redshift--radio power degeneracy to be broken.  Partial rank
correlations and principal component analysis have been used to
determine which of redshift and radio power are the critical
parameters underlying the observed variation of the ionization state
and kinematics of the emission line gas. 

\oohb is shown to be a useful ionization mechanism diagnostic. 
Statistical analysis of the data shows that the ionization state of
the emission line gas is strongly correlated with radio power, once
the effects of other parameters are removed. No dependence of
ionization state on cosmic epoch is observed, implying that the
ionization state of the emission line gas is solely a function of the
AGN properties rather than the host galaxy and/or environment.

Statistical analysis of the kinematic properties of the emission line
gas shows that these are strongly correlated independently with both
redshift and radio power.  The correlation with redshift is the
stronger of the two, suggesting that host galaxy composition or
environment may play a role in producing the less extreme gas
kinematics observed in the emission line regions of low redshift
galaxies.  

For both the ionization and kinematic properties of the galaxies, the
independent correlations observed with radio size are stronger than
with either radio power or redshift.  Radio source age is clearly a
determining factor for the kinematics and ionization state of the
extended emission line regions.
\end{abstract}


\section{Introduction}
\footnotetext{E-mail: kji@mrao.cam.ac.uk}

Powerful radio sources are typically associated with massive
ellipticals, and are often observed to have extended emission line
regions aligned along the radio source axis.  The luminosity and
kinematic properties of the emission line regions are generally 
observed to be more extreme for radio sources at high rather than at
low redshifts.  
It is important to establish exactly how the
properties of the extended emission line regions (ionization
mechanism, kinematics and physical extent) are functions of redshift,
radio power and radio size.  This will add to our understanding of
these complex systems.

Baum \& McCarthy (2000) carried out a spectroscopic study of a sample
of 52 radio galaxies covering a large range of redshifts, most of
which are selected from the 3CR sample. Their analysis of the
kinematic and morphological properties of these objects demonstrated a
number of important correlations.  The kinematic properties of the
emission line gas in these sources were found to vary strongly with
redshift and/or radio power, the higher redshift galaxies generally
displaying greater line widths and velocity amplitudes.  The inferred
mass of ionized gas and the enclosed dynamical gas were also seen to
increase with redshift.  Although jet-cloud interactions are not
excluded by the data, Baum \& McCarthy preferred a gravitational
origin for the observed kinematics.   

Best, R\"{o}ttgering \& Longair (2000a,b) carried out a study of the 
emission line regions of high redshift 3CR galaxies at $z \sim 1$.
Small radio sources (those with a radio size $D_{rad} <
120\,\rm{kpc}$) were observed to exist in a lower ionization state,
and to possess more distorted velocity profiles and boosted
low-ionization (e.g. [O\textsc{ii}]) emission
line luminosities in comparision with larger radio galaxies ($D_{rad}
> 120\,\rm{kpc}$).  Their spectra were consistent with small radio
sources being predominantly shock ionized, and large radio sources
photoionized by the AGN.  Emission line regions were generally larger
in spatial extent in the smaller radio sources.  The kinematic and
ionization properties of these emission line regions were also found
to be strongly correlated.

Best {\it et al} also compared the properties of the emission line
regions of high redshift 3CR galaxies at $z \sim 1$ with a sample of
low redshift 3CR galaxies with z $\lta 0.2$ (from Baum {\it et al}
1992).   Similar radio size trends were observed in the low redshift
sample, although some trends with either radio luminosity and/or
redshift were also observed.  Line luminosities and equivalent widths
were generally less extreme in the low redshift sources.  Tadhunter
{\it et al} (1998) showed that a similar correlation exists between
line luminosity and radio luminosity in a subsample of 2\,Jy sources
from the sample of Wall \& Peacock (1985).   Assuming a simple quasar
illumination model, the observed ionization state of the emission line
gas implied that this correlation could not be explained by a simple
increase in the flux of ionizing photons associated with more powerful
AGN at higher redshifts.  The observations could however be reconciled
with the simple photoionization model if emission line region cloud
density is enhanced at higher redshifts, either due to changes in the
host galaxy environment or radio source shocks increasing in importance.

The flux--limited 3CR sample suffers, however, from Malmquist bias:
the strong correlation between redshift and radio power in a flux
limited sample prevents the effects of changes in these two parameters
from being disentangled from each other.  This degeneracy needs to be
broken, in order to determine to what extent changes in redshift
and/or the host galaxy environment, in particular any possible
variations in the size and structure of the emission line regions and
the density of the IGM, influence the ionization and kinematics of the
extended emission line regions of these galaxies.   
Higher radio power galaxies generally have intrinsically more powerful
jets; the bulk kinetic power of the radio jets has been shown to be
strongly correlated with narrow line region luminosities (Rawlings \&
Saunders 1991).  It is important that we fully understand the effects
of changes in radio power on the ionization state and kinematic
properties of the extended emission line regions (EELRs).

Over recent years we have been involved in an ongoing programme of
multiwavelength observations investigating a subsample of 11 6C radio
galaxies, (Eales 1985; Best {\it et al} 1999; Inskip {\it et al
(hereafter Paper 1)}). These sources were selected in order to be well
matched to a previously well studied subsample of 28 3CR galaxies with
redshift $z \sim 1$, (Longair, Best \& R\"{o}ttgering 1995; Best,
Longair \& R\"{o}ttgering 1996, 1997; Best {\it et al} 1998, 2000a,b).
In Paper 1 the results of deep spectroscopic observations of eight of
these sources using the William Herschel Telescope (WHT) were
presented.  We refer the reader to Paper 1 for details of sample
selection, observations and data reduction, reduced 1- and
2-dimensional spectra, tabulated line fluxes and composite spectra.
Also included in Paper 1 is a discussion of the kinematic and
ionization properties of the 6C galaxies, contrasting them with those
of the spectroscopic sample of 3CR galaxies at $z \sim 1$. 

The major results of Paper 1 are briefly summarized below:
\vspace{-10pt}
\begin{enumerate}
\item The kinematical properties of the EELRs of 6C galaxies are
similar to those of the more powerful 3CR sources studied at the same
redshift.  Small radio sources generally possess more extensive
emission line regions with a more distorted velocity profile and more
extreme kinematics than those of larger radio sources.   
\item \oo emission line luminosity is anticorrelated with the size (or
age) of the radio source. 
\item Ionization state varies similarly with radio size for both
subsamples, despite the decrease in radio power of the 6C sources.
This is interpreted as being due to a changing contribution of
ionizing photons from the shock front associated with the expanding
radio source.  The optical/UV spectra of the EELRs associated with
large radio sources are dominated by photoionization by the AGN and
those of small sources by shocks. 
\end{enumerate}

\vspace{-7pt}
There is a great deal of evidence that shocks associated with the
expanding radio source play a major role in creating the extreme gas
kinematics observed in the spectra of radio galaxies. In addition to
the results outlined above, the extreme line widths observed in the
extended emission line regions are coincident with the radio source
structures.   This argues against the gravitational origin for the
observed kinematics of the gas preferred by Baum \& McCarthy, at least
for the more extreme sources.

In this paper we compare the results of the 3CR and 6C $\sim 1$
subsamples with lower redshift 3CR sources, selected to have similar
radio powers to those in the 6C subsample.  By considering samples of
galaxies covering a larger region of the $P$--$z$ plane (where $P$ is
the radio luminosity of the galaxies) we are able to
break the degeneracy between redshift and radio power.  

The structure of the paper is as follows.  In section 2, we discuss the
selection of low redshift galaxies for comparison with the $z \sim 1$
3CR and 6C subsamples. The results of our comparison between low and
high redshift sources are presented in section 3, and discussed in
section 4. Conclusions are drawn in section 5.  Values for the
cosmological parameters $\Omega_0=0.3$, $\Omega_\Lambda=0.7$ and
$H_{0}=65\,\rm{km\,s^{-1}\,Mpc^{-1}}$ are assumed.

\section{selection of the low redshift sample}

The comparison of the 6C spectroscopic data with those of a more
powerful sample of 3CR galaxies at the same redshift gives us an
insight into the variation of the ionization states and kinematical
properties of radio galaxies with radio power.  By selecting
another subsample of galaxies at lower redshift, matched in radio
power to the 6C subsample at $z \sim 1$, we can investigate the
dependence of these galaxy properties with redshift alone, as the
combination of three such samples breaks the $P$--$z$ degeneracy.
Unfortunately there are no data available in the literature exactly
matching our criteria of full ionization and kinematic data for a
large number of galaxies matched in radio power to the 6C
subsample. The low redshift data used in this paper has therefore been
taken from two different samples: ionization data from the sample of
Tadhunter {\it et al} (1998), and the low redshift kinematic data from
the sample of Baum \& McCarthy (2000). 

For ionization studies we have selected narrow line radio galaxies
from the steep spectrum selected subsample of Tadhunter {\it et al}
(1993), which is itself taken from the 2\,Jy sample of Wall \& Peacock
(1985).  The Tadhunter sample includes all sources from the Wall \&
Peacock sample with 2.7\,GHz flux density $> 2$Jy, declinations
$\delta < 10^\circ$ and redshifts $z < 0.7$. Whilst the full 2-Jy
sample is not spectroscopically complete, the remaining objects
without identified redshifts are very faint, and unlikely to lie
within the $z < 0.7$ subsample of Tadhunter {\it et al} (1993).   This
upper redshift limit is also the lower redshift limit of the
spectroscopic subsample of 3CR galaxies of Best {\it et al} (2000a),
which spanned the redshift range $0.7 < z < 1.25$.  The eighteen
sources selected have both [O\textsc{iii}]/[O\textsc{ii}] and
[O\textsc{iii}]/H$\beta$ line ratios available, as presented by
Tadhunter {\it et al} (1998).  Eight of these sources lie in a similar
radio luminosity range to the 6C $z \sim 1$ subsample, and thus can be
used in a statistical comparison with the high redshift sources.  Two
sources were excluded on the basis of incomplete emission line
information, but neither of these would have fallen in the luminosity
range selected for comparison to the 6C galaxies.  The sources
included in the statistical comparison subsample are 0023-26, 0039-44,
0859-25, 1306-09, 1934-63, 2250-41, 2314+03 (3C459) and 2356-61. 
The full subsample
of eighteen galaxies is 90\% complete, and unbiased. Although this
sample provides ionization data suitable for comparison with the $z
\sim 1$ subsamples, we need to look elsewhere for the kinematics of
low redshift radio sources.  
 
Baum \& McCarthy (2000) provide an analysis of the kinematical and
morphological properties of a sample of 52 radio galaxies. Their
sample is a compilation of two other samples: low redshift galaxies
are taken from the samples of Baum, Heckman \& van Breugel (1990, 1992)
and the intermediate--high redshift galaxies from the sample of
McCarthy {\it et al} (1996).  Both of these were drawn in turn from
the emission line imaging survey of Baum {\it et al} (1988) and
imaging by McCarthy {\it et al} (1995).  The Baum \& McCarthy sample
is slightly biased against sources with small emission line regions,
particularly at higher redshifts; at redshifts greater than $\sim
0.2$, it includes most 3CR sources with emission line imaging sizes
greater than 5\arcsec.  For comparison with the kinematical properties
of $z \sim 1$ radio galaxies, we have selected twenty-two FRII
galaxies from the sample of Baum \& McCarthy (2000).  These are all
FRII sources with redshifts of $z < 0.7$, the same upper redshift
limit as used in the Tadhunter {\it et al} subsample.   Of the
twenty-two galaxies selected, twenty-one have a measured FWHM, but
only eight are used in statistical 
comparisons with the 6C galaxies at $z \sim 1$, as these have similar
radio luminosities.   These are the sources 3C79, 3C169.1, 3C299, 3C300,
3C306.1, 3C337, 3C435 and 3C458.

\begin{figure}
\vspace{2.1 in}
\begin{center}
\includegraphics{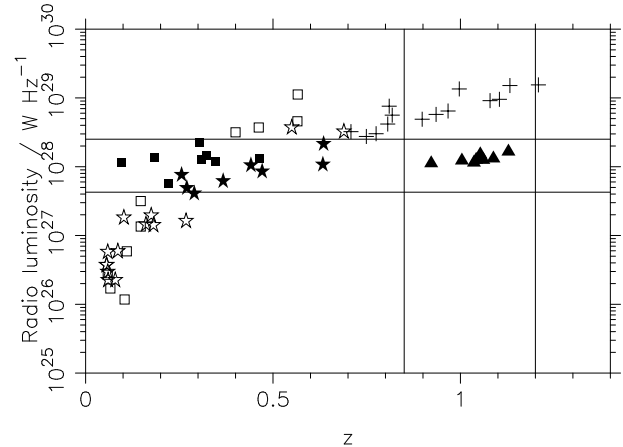}
\end{center}
\caption{$P$--$z$ diagram illustrating the selection of low redshift
galaxies. The 3CR $z \sim 1$ subsample radio galaxies are represented
by crosses, and the 6C radio galaxies by solid triangles.  The
vertical lines delineate the redshift range from which the 6C
subsample was selected, and the horizontal lines the range in radio
luminosity within which low redshift sources will be directly compared
with 6C sources. The Baum \& McCarthy low redshift subsample radio
galaxies are represented by filled and open stars, and the Tadhunter 
{\it et al} sample radio galaxies by filled and open squares, the
filled symbols representing the luminosity--matched low redshift data
used in the statistical comparisons.
\label{Fig: 1}}
\end{figure}

Fig.~\ref{Fig: 1} shows the radio--power vs. redshift distribution of
the four samples considered in this paper, illustrating which of the
low redshift galaxies are suitable for comparison with the $z \sim 1$
6C and 3CR subsamples. 

\section{Comparison with galaxies at low redshift}
\subsection{The ionization state of the emission line gas}
\subsubsection{[O\textsc{ii}]/H$\beta$ as a diagnostic test}
 
\begin{figure*}
\vspace{4.7 in}
\begin{center}
\includegraphics{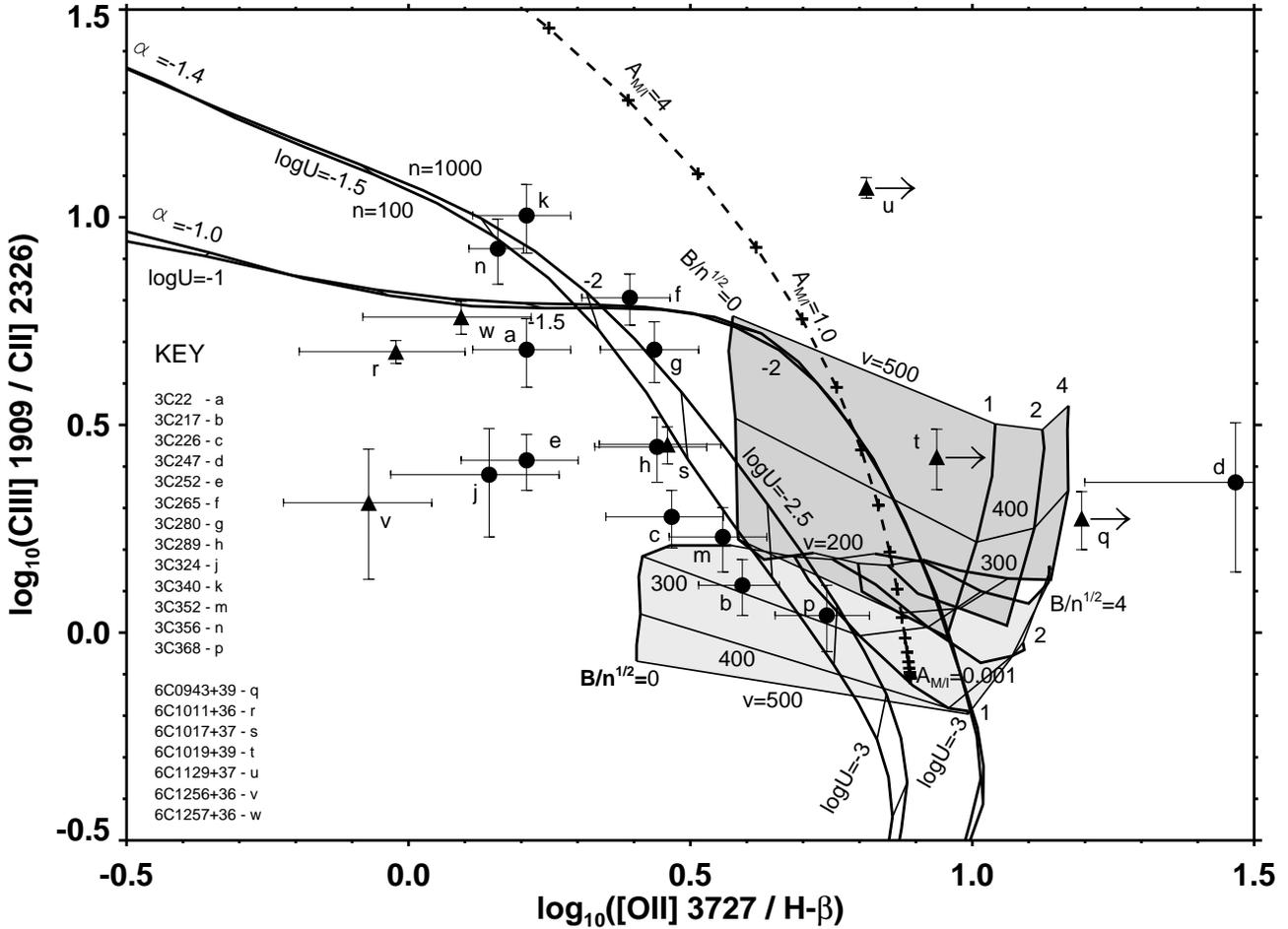}
\end{center}
\caption{An \oohb vs \crat emission line ratio diagnostic diagram for
6C and 3CR galaxies at $z \sim 1$.  Points are plotted for each
galaxy (3CR galaxies are represented by circles and 6C galaxies by
triangles), and compared with the theoretical predictions for shock
ionization, simple photoionization, and photoionization including
matter-bounded clouds.  The shock ionization line ratio predictions
are those of the models of Dopita and Sutherland (1996). Results for
both simple shock ionization (light shading) and also for models
including a precursor ionization region (dark shading) are included 
on the diagram. Shock velocities were allowed to vary between 150 to
500\,km\,s$^{-1}$. The `magnetic parameter', $B/\sqrt{n}$, which controls
the effective ionization parameter of the post-shock gas, was varied
from 0 to $4\,\mu$G\,cm$^{-1.5}$. The simple photoionization model
tracks are taken from the theoretical line ratios calculated using the
\textsc{mappings ii} code (Sutherland, Bicknell \& Dopita 1993, Allen
{\it et al} 1998, Mark Allen {\it private communication}).   A power
law spectrum illumination ($F_{\nu} 
\propto \nu^{\alpha}$, $\alpha = -1$ or $-1.4$, with a high energy
cut-off at 1.36\,keV) of a planar slab of material ($n_e = 100$
or $1000\,\rm{cm}^{-3}$) was assumed, with an ionization parameter
$10^{-4}\,\leq\,U\,\leq\,1$. The models correspond to cloud sizes
from 0.003 to 32 parsec, and are ionization bounded. Photoionization
tracks for the combination of matter bounded and ionization bounded
clouds are also plotted, using the predicted line ratios of Binette
{\it et al} (1996).  $A_{M/I}$ represents the ratio of the solid angle
from the photoionizing source subtended by matter bounded clouds
relative to that of ionization bounded clouds.  For further details,
see Best {\it et al} 2000b
\label{Fig: 2}}
\end{figure*}

For the Tadhunter {\it et al} subsample, the line ratios
[O\textsc{iii}]/[O\textsc{ii}] and [O\textsc{iii}]/H$\beta$ are
available. These can be used to determine [O\textsc{ii}]/H$\beta$, a
line ratio which can be inferred for the $z \sim 1$ data based on our
measurements of [O\textsc{ii}] and H$\gamma$ or other Balmer lines,
assuming the Balmer line ratios for case B recombination.  The
composite spectra produced for the $z \sim 1$ samples (Paper 1)
suggest that this line ratio is quite different for large and small
radio sources, and thus may be a useful diagnostic of the ionization
mechanism.  To investigate this, we used the \textsc{mappingsii} 
(Sutherland, Bicknell \& Dopita 1993) 
results for \oohb and \crat to create a new ionization mechanism
diagnostic diagram (Fig.~\ref{Fig: 2}).   

Marked on the diagram are the predictions of shock (Dopita \&
Sutherland 1996) and photoionization models (Binette, Wilson \&
Storchi-Bergmann 1996; Allen, Dopita \& Tsvetanov 1998; digitised data
kindly provided by Mark Allen).  The 6C and 3CR data points are also
included.  Where H$\gamma$, H$\delta$ or H$\zeta$ could not be
measured in our spectra, an upper limit was determined for the
H$\beta$ flux for these sources.  In these cases, the \oohb line ratio
is given as a lower limit.  No correction has been made for internal
extinction.  Assuming similar levels of internal extinction to that
measured for Cygnus A (Osterbrock, 1989), this is likely to have only
a minimal impact on our results.   

The diagnostic plot in Paper 1 utilised the predictions of the line
ratios \crat and \nrat for these models.  Large sources were generally
well matched to pure photoionization by the AGN, with an ionization
parameter $-2 \lta \rm{log_{10}U} \lta -1$ and spectral index $\alpha =
-1$.  Small sources fitted well the predictions of the shock models
incorporating precursor ionization regions.  Alternatively, the
photoionization models combining matter bounded and ionization bounded
clouds of Binette {\it et al} also seemed to provide a good fit to the
data, with the smaller radio sources typically having a smaller
fraction of matter bounded clouds.  The positions of the data points
in relation to the different models can be compared between
Fig.~\ref{Fig: 2} and the diagnostic plot from Paper 1, allowing the
consistency of the models in fitting the observed data to be gauged.   

When comparing two or more line ratio diagnostic diagrams, it quickly
becomes apparent that not all the models explain the spectra of the
observed sources consistently from one plot to another.  However,
some of the modelled tracks appear to fit the data reasonably well on
both plots.  The five `photoionized' sources on the diagnostic plot in
Paper 1 appear close to the predicted line ratios of the same model in
Fig.~\ref{Fig: 2}. Most of the points which were previously well
matched by the predictions of the shock plus precursor region models
remain so in this diagram as well.   As a general rule, small sources
show lower ionization parameters than larger sources, and their
positions on the diagnostic plot are consistent with the predictions
of shock ionization.  However, the shock ionization and AGN
photoionization models cannot be completely separated in
Fig.~\ref{Fig: 2}.  These results clearly illustrate that the power of
line ratio plots as a diagnostic of the ionization mechanism is
greatly enhanced when several different plots are considered together,
as ambiguities on one or more plots may be resolved by additional
modelled line ratios. 

One clear result from this plot is that the preferred region of the
photionization tracks predicts lower values of \oohb than the
predictions of the shock models. Thus, the \oohb line ratio 
can provide a suitable indication of any changes of ionization state which
may be observed in the low redshift sources, just as the \crat line
ratio was used to do so in Paper 1.

\subsubsection{Balmer lines vs. [OII]3727 - investigating the
variation in ionization state within the three samples.}

\begin{figure}
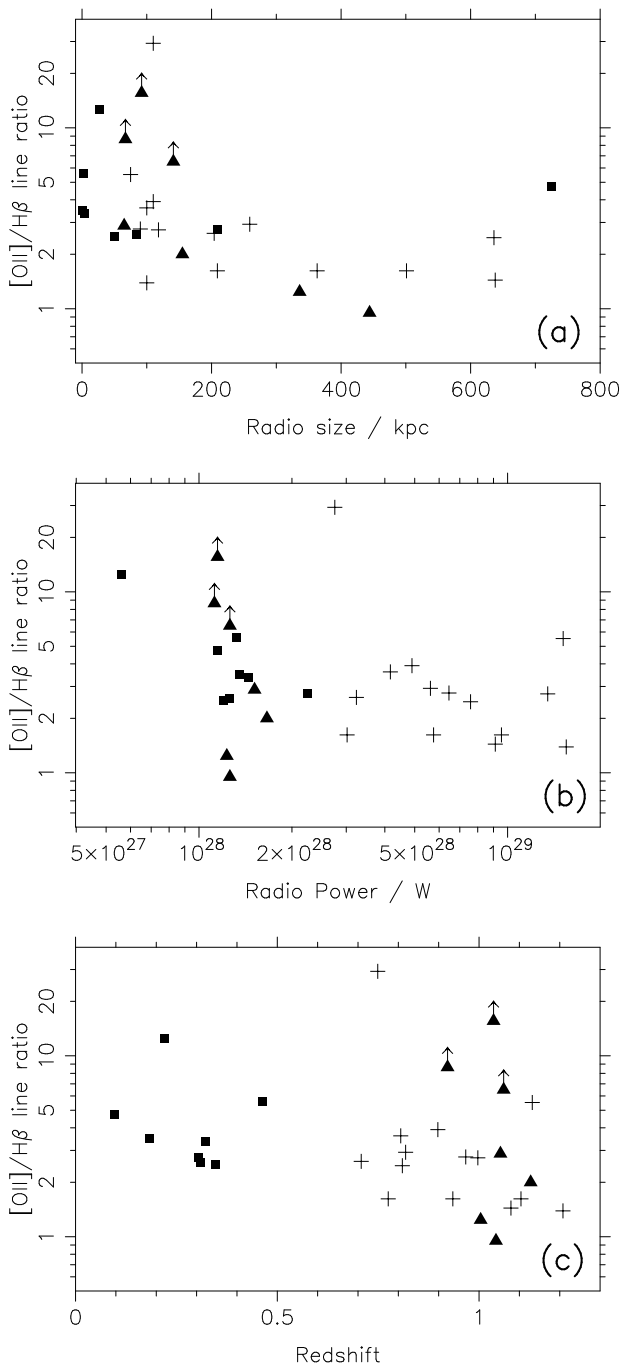

\vspace{7.0 in}
\begin{center}
\includegraphics{lowz_O2Hb_Drad.ps}
\includegraphics{lowz_O2Hb_Lrad.ps}
\includegraphics{lowz_O2Hb_z.ps}
\end{center}
\caption{Plot of the [O\textsc{ii}]/H$\beta$ line ratio vs. (a) -- Radio
size, (b) -- Radio power and (c) redshift for three samples. Crosses
represent the 3CR $z \sim 1$ subsample galaxies, filled triangles the
6C sources, and the filled squares the Tadhunter {\it et al} sample
galaxies matched in radio power to the 6C subsample. The 6C and 3CR
data points are plotted as lower limits where 
the H$\beta$ luminosities are upper limits inferred from H$\delta$
rather than H$\gamma$ luminosities. 
\label{Fig: 3}}
\end{figure}

In Fig.~\ref{Fig: 3}a the [O\textsc{ii}]/H$\beta$ line ratio data for
the Tadhunter {\it et al} subsample and the two $z \sim 1$ samples are
plotted against radio size.  Fig.~\ref{Fig: 3}b displays the variation
of this line ratio with radio power for the same samples;
Fig.~\ref{Fig: 3}c plots this parameter against redshift.  It is
interesting to note that there is a much larger spread in the
[O\textsc{ii}]/H$\beta$ line ratio for the lower radio power sources. 

Spearman rank correlation tests have been carried out for these data 
and the results tabulated in Table 1.  For three 6C sources, the \oohb
line ratio could only be calculated as a lower limit; these values
have been treated as measured values for all statistical analyses 
unless stated otherwise.  However, as the \oohb line ratios for these
sources are amongst the highest in the samples, the statistics are
unlikely to suffer.  The line ratios of the galaxies in all three
subsamples combined are anticorrelated with radio size at the 99.8\%
significance level, suggesting that this line ratio is a suitable
alternative to \crat as an indicator of changes in ionization state
with radio size.   The data are slightly less strongly correlated with
radio power and redshift.  

We have also carried out partial rank correlations for these galaxies,
using the method described by Macklin (1982), which enables the true
dependencies of the different properties to be determined, by removing
the effects of correlations with one or more of any other parameters.   

Our results show strong anticorrelations of the \oohb line
ratio with both radio size and luminosity (see Table 1). Although a
Spearman rank test initially found a significant correlation with
redshift, partial rank analysis shows that redshift is an unimportant 
parameter and the correlation was induced by the strong correlation
between $z$ and radio power.  We can conclude that there is no strong
dependence of ionization state on any changes with redshift in the
properties of the host galaxy or its environment.  

Another way of determining the most important correlations within a
set of data is a principal component analysis (also known as a
Karhunen-Loeve transform).  This is described by Kendall (1980) and
Efstathiou \& Fall (1984).  For a data set with {\it p} variables for
each of {\it N} parameters, a new set of uncorrelated parameters can
be determined (the principal components of the data) which are linear
functions of the original variables.  If the original variables are
standardized to zero mean and unit variance, the principal components
will also be independent of each other.  The first principal component
of the data has the largest possible variance of any linear function,
and can be thought of as defining the line of closest fit to
the {\it p}--dimensional set of {\it N} data points, in that the sum
of squares of the distances of the data points in any direction from
such a line will be minimised.  The 
magnitude of this eigenvector indicates what fraction of the dataset
can be explained by these correlations alone.  The second principal
component has the largest variance subject to being uncorrelated
(i.e. orthogonal) to the first. Further principal components are
uncorrelated with all the preceding principal components.

The results of a principal component analysis of our data are
displayed in Table 2, and confirm the results of the partial rank
coefficients.  This analysis has been carried out twice for the three
galaxy subsamples; first using only the low-redshift data matched in
radio power to the 6C subsample, and secondly using the full low
redshift subsample.  Although there are subtle differences between the
two sets of results, the overall correlations determined are the same
in both cases. 

In both cases, the eigenvector of the first principal component
accounts for $\sim 45$\% of the variance within the sample, that of
the second component $\sim 30\%$, the third component $\sim 20\%$ and
the fourth $<10\%$.  A positive correlation between $P$ and $z$ is
clearly the strongest correlation within the data, being inverted
significantly only in the final eigenvector, in order to remove the
remaining scatter in the data.  The advantage of using the chosen
combination of three samples is that the effects of this correlation
can be accounted for, and the redshift--radio power degeneracy
successfully broken.   Anti-correlations of the \oohb line ratio with
both redshift and radio power are implied by the first principal
component.  However the second component reverses and weakens these
correlations, particularly for redshift.  In the first eigenvector,
ionization state and radio size ($D_{rad}$) are seen to be
anticorrelated for the matched samples, and very weakly correlated for
the full low redshift sample.  The second eigenvector shows a strong
anticorrelation between the ionization state and size of the radio
source in both cases, suggesting that this is indeed a significant
correlation.  This difference is perhaps due to an underlying
anticorrelation between radio size and either $P$ or $z$, which
weakens the relationship between radio size and the \oohb line ratio,
although it is unclear from these data whether anticorrelations
between either $P$ and $D_{rad}$ or between $z$ and $D_{rad}$ are
real.   An anticorrelation between $D_{rad}$ and $z$ is observed by
Neeser {\it et al} (1995) and Blundell {\it et al} (1999), and is
likely to be due in part to the selection effects inherent within flux
limited samples, which are biased towards smaller sources at high
redshift due to Malmquist bias and the decrease in flux density with
radio size for individual sources (Kaiser, Dennett--Thorpe \&
Alexander 1997).  

\begin{table*}
\caption{Spearman Rank Correlation Coefficients. These are given for
the correlations of the emission line ratio \oohb with
redshift, radio power and radio size, along with 
partial rank coefficients for the same parameters. Significance levels
for each correlation are also tabulated where these are greater than
80\%. I = ionization as indicated by the line ratio [O\textsc{ii}]/H$\beta$, z =
redshift, P = radio power and D = radio size.  The labelling of the
correlations is such that $\rm{r}_{\rm{Iz}}$ is the correlation
coefficient for ionization state versus redshift, and
$\rm{r}_{\rm{Iz|D}}$ is the partial correlation coefficient, removing
correlations with radio size from any correlation between ionization
state and redshift.  $\rm{r}_{\rm{Iz|PD}}$ is the partial correlation
coefficient for ionization state and redshift after correlations with
both radio size and radio power have been removed.  Other correlation
coefficients are labelled similarly.} 
\begin{center} 
\begin{tabular}{cc@{=}r@{.}lc@{=}r@{.}lc@{=}r@{.}l} 
\hline
&\multicolumn{9}{c}{Ionization correlations...}\\
&\multicolumn{3}{c}{...with redshift}&\multicolumn{3}{c}{...with radio power}&\multicolumn{3}{c}{...with radio size}\\\hline
3C at $z \sim 1$ ($n$=14)                   &$\rm{r}_{\rm{Iz}}$&-0&338 (-88\%)&$\rm{r}_{\rm{IP}}$   &-0&369 (-90\%)&$\rm{r}_{\rm{ID}}$   &-0&555 (-98\%)\\
6C at $z \sim 1$ ($n$=7)                    &$\rm{r}_{\rm{Iz}}$&-0&214        &$\rm{r}_{\rm{IP}}$   &-0&505 (-88\%)&$\rm{r}_{\rm{ID}}$   &-0&750 (-97\%)\\ 
Radio power matched low redshift 3C ($n$=8) &$\rm{r}_{\rm{Iz}}$&-0&310        &$\rm{r}_{\rm{IP}}$   &-0&333        &$\rm{r}_{\rm{ID}}$   &-0&310 \\\\
All three subsamples                        &$\rm{r}_{\rm{Iz}}$&-0&332 (-96\%)&$\rm{r}_{\rm{IP}}$   &-0&369 (-98\%)&$\rm{r}_{\rm{ID}}$   &-0&525 (-99.8\%)\\
($n$=29)                                    &$\rm{r}_{\rm{Iz|P}}$& 0&053      &$\rm{r}_{\rm{IP|D}}$ &-0&202 (-85\%)&$\rm{r}_{\rm{ID|z}}$ &-0&256 (-91\%)\\
                                            &$\rm{r}_{\rm{Iz|D}}$&-0&065      &$\rm{r}_{\rm{IP|z}}$ &-0&238 (-89\%)&$\rm{r}_{\rm{ID|P}}$ &-0&220 (-87\%)\\
                                            &$\rm{r}_{\rm{Iz|PD}}$& 0&051     &$\rm{r}_{\rm{IP|Dz}}$&-0&198 (-84\%)&$\rm{r}_{\rm{ID|zP}}$&-0&219 (-86\%)\\\\

Including the full low-z subsample          &$\rm{r}_{\rm{Iz}}$&-0&247 (-93\%)&$\rm{r}_{\rm{IP}}$   &-0&256 (-94\%)&$\rm{r}_{\rm{ID}}$   &-0&388 (-99.2\%)\\
($n$=38)                                    &$\rm{r}_{\rm{Iz|P}}$& 0&046      &$\rm{r}_{\rm{IP|D}}$ &-0&194 (-87\%)&$\rm{r}_{\rm{ID|z}}$ &-0&196 (-88\%)\\
                                            &$\rm{r}_{\rm{Iz|D}}$&-0&076      &$\rm{r}_{\rm{IP|z}}$ &-0&139        &$\rm{r}_{\rm{ID|P}}$ &-0&230 (-91\%)\\
                                            &$\rm{r}_{\rm{Iz|PD}}$& 0&072     &$\rm{r}_{\rm{IP|Dz}}$&-0&192 (-87\%)&$\rm{r}_{\rm{ID|zP}}$&-0&236 (-92\%)\\
\hline
\end{tabular}
\end{center}
\end{table*}

\begin{table*}
\caption{Principal component analysis of emission line region
ionization state and other radio source parameters. This analysis has
been carried out on the two $z \sim 1$ subsamples and either the radio
power matched or full low-redshift subsamples.  The four principal
components are linear functions of the parameters, determined such
that the first component defines the closest line of fit to the data
points.  The eigenvectors defining each principal component are given,
as are the percentages of the total variance in the data accounted for
by each principal component.} 
\begin{center} 
\begin{tabular} {ccr@{.}lr@{.}lr@{.}lr@{.}lcccr@{.}lr@{.}lr@{.}lr@{.}lc} 
\hline
{Variable} & \multicolumn{10}{c}{Eigenvectors for the three matched
samples} && \multicolumn{10}{c}{Eigenvectors including the full
low redshift data}\\ 
&&\multicolumn{2}{c}{1}&\multicolumn{2}{c}{2}&\multicolumn{2}{c}{3}&\multicolumn{2}{c}{4}&&&&\multicolumn{2}{c}{1}&  \multicolumn{2}{c}{2}&\multicolumn{2}{c}{3}&\multicolumn{2}{c}{4}\vspace{2.5pt}\\
I&& 0&4012 & 0&5745 & 0&6905 & 0&1794 &&&& 0&1336 & 0&7273 & 0&6572 & 0&1460& \\
z&&-0&5413 & 0&5283 & 0&0446 &-0&6526 &&&&-0&6650 &-0&0685 & 0&3561 &-0&6530& \\
P&&-0&6238 & 0&2694 &-0&0519 & 0&7319 &&&&-0&6972 &-0&0225 & 0&0074 & 0&7165& \\
D&&-0&3961 &-0&5642 & 0&7201 &-0&0790 &&&& 0&2320 &-0&6826 & 0&6643 & 0&1975& \vspace{2.5pt}\\
{Percentage} && \multicolumn{2}{c}{44.30}& \multicolumn{2}{c}{26.43}&\multicolumn{2}{c}{18.45}&\multicolumn{2}{c}{10.82}&&&& \multicolumn{2}{c}{43.28} &\multicolumn{2}{c} {29.74} & \multicolumn{2}{c}{19.30} & \multicolumn{2}{c}{7.68}&\\
\hline
\end{tabular}
\end{center}
\end{table*}

\begin{figure}
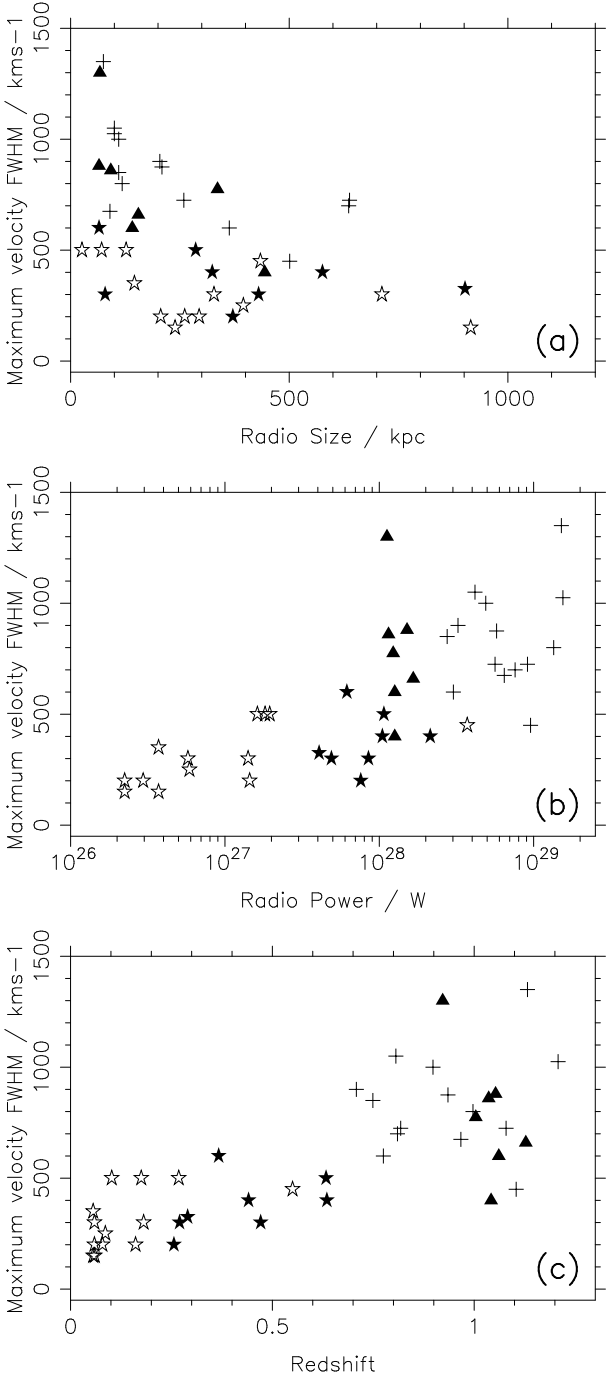

\vspace{7.0 in}
\begin{center}
\includegraphics{lowz_FWHM_Drad.ps}
\includegraphics{lowz_FWHM_Lrad.ps}
\includegraphics{lowz_FWHM_z.ps}
\end{center}
\caption{FWHM of galaxies in three samples. (a) -- top: FWHM vs. Radio size,
(b) -- middle: FWHM vs. Radio power, (c) -- bottom: FWHM vs. Redshift.
The 3CR $z \sim 1$ subsample galaxies are represented by crosses, and
the 6C galaxies by solid triangles. The Baum \& McCarthy low redshift subsample
galaxies are represented by filled and open stars, for those galaxies matched in flux to 6C
sources and the remainder of the low redshift sources respectively.
\label{Fig: 4}}
\end{figure}

These results are fully consistent with the results of the partial
rank correlation tests, and suggest that the following are real: a
strong positive correlation between $P$ and $z$, and weaker
anticorrelations between $D_{rad}$ and $z$ or $P$, due to the selection
effects in defining the sample.  Anticorrelations
between $I$ and $P$, and between $I$ and $D_{rad}$ are also real,
where `$I$' represents the \oohb line ratio as a measure of the
ionization state within the emission line gas, a large value for $I$
implying a low ionization state.  Of the implied correlations with
ionization state, the \oohb line ratio is most strongly anticorrelated
with radio size, followed by radio power.  

\subsection{The Baum \& McCarthy sample: Kinematics comparison}

The first parameter we consider in our comparison of the kinematics of
radio galaxies at high and low redshift is the maximum value of the
velocity FWHM, which varies with spatial position.  This parameter is
very robust, easily determined and provides a straightforward
measure of the kinematic effects of any shocks associated with
jet-cloud interactions.  The emission line gas FWHM is plotted
against radio size, radio power and redshift in Fig.~\ref{Fig: 4}.  

In general, the lower redshift sources have smaller FWHM than the two 
high redshift samples.  The FWHM of both 6C and 3CR galaxies are
anticorrelated with radio size, whilst the low redshift galaxies
selected from the Baum \& McCarthy sample are not correlated very
significantly with this parameter.  Baum \& McCarthy claim that there 
is little evidence for an anticorrelation between radio size and the
velocity FWHM of the emission line gas in their complete sample, such
as that seen in both $z \sim 1$ subsamples. It is not surprising that
this result is obtained, as by considering galaxies over their full
range of redshifts, any positive correlation between redshift and FWHM
such as that observed for our three samples will act to diminish any
anticorrelation between radio size and FWHM.  Best {\it et al} (2001)
separated the full Baum \& McCarthy sample into subsamples of sources
with $z < 0.6$ and $z > 0.6$; the high and low redshift subsamples
each displayed a strong anticorrelation between velocity FWHM and
radio size.  In the redshift range we consider, the sample is nearly
complete for sources with large emission line regions and an
anticorrelation between FWHM and redshift is observed.  Considering
all three samples together, FWHM is clearly anticorrelated with radio
size, and correlated with redshift and radio power. The strengths of
these correlations are listed in Table 3. 

As before, some of these correlations may not indicate an intrinsic
connection between FWHM and redshift, radio power or radio size, but
instead be due to secondary correlations between these other parameters.
This can be disentangled  by carrying out partial rank
correlation tests on the data, removing the influence of one or
more of the other parameters.  The partial rank analysis shows that
the strongest relationship between the parameters is an anticorrelation
with radio size ($> 99.9$\% significant), followed by a positive
correlation with redshift ($> 98$\% significant) and a slightly
weaker correlation with radio power ($> 94$\% significant).  The
relative importance of the trends with redshift and radio power can
also be observed in Fig.~\ref{Fig: 4}.  Fig.~\ref{Fig: 4}b shows that
the kinematic properties of the 6C subsample are clearly more extreme
than those of the low redshift 3CR subsample.  These samples are
matched in radio power, but not in redshift.  The differences between
the $z \sim 1$ 6C and 3CR subsamples, matched in redshift but not
radio power, are much less obvious, suggesting that the gas kinematics
are indeed more strongly dependent on redshift than on radio power. 

We have also carried out a principle component analysis for these data
sets, the results of which are tabulated in Table 4.  The first
principal component accounts for over 60\% of the variance in the
kinematic data. Practically all the scatter in the data can be
accounted for with the first two eigenvectors.  The reduction in
the scatter of the data points compared with the ionization state data
provides the first indication that any correlation between $P$, $D$ or
$z$ with the emission region gas FWHM is likely to be much more
significant than the correlations observed between these parameters
and the ionization state of the gas.  The first eigenvector for both
data sets gives strong positive correlations between $P$, $z$ and
FWHM, and a strong anticorrelation between each of these parameters 
and $D_{rad}$.  In the second eigenvector the correlation between $P$
and $z$, and the anticorrelation between FWHM and $D_{rad}$ remain
strong, but the 
other correlations implied by the first eigenvector are reversed and
hence greatly weakened (particularly the FWHM/radio power
correlation).  These results fully confirm those of
the partial rank analysis: the most important kinematic correlations 
are FWHM anticorrelated with radio size, and positively correlated
with redshift and radio power, in decreasing order of significance.
As for the ionization data we also observe a correlation of redshift
with radio power, as well as an anticorrelation of radio size with
redshift.  
 
\begin{figure}
\vspace{2.1 in}
\begin{center}
\includegraphics{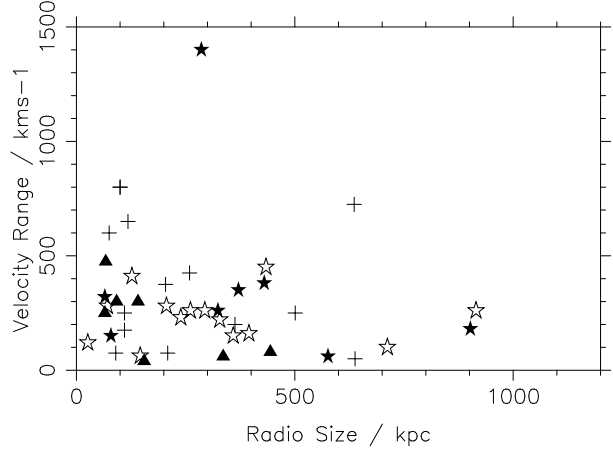}
\end{center}
\caption{Variation in velocity range with radio size for three
samples.  Symbols as in Figure 4.
\label{Fig: 5}}
\end{figure}

The variation of velocity range (defined as the difference between the
most positive and negative velocity components of the \oo line) with
radio size is plotted in Fig.~\ref{Fig: 5} for all three samples. 
The distribution of all three samples appears very similar, suggesting
that velocity range has no dependence on either
redshift or radio power.  A visual examination of Fig.~\ref{Fig: 5}
suggests that the greatest range of velocities is observed in the
smaller radio galaxies.  However, a Kolmogorov-Smirnov test, which is
more suitable than a Spearman rank test for an analysis of the
significance of the suspected trend, shows that the decrease in
velocity range with radio size is not significant for any or all of
the subsamples combined; nor is the decrease in maximum velocity range
from the 3C to 6C sources statistically significant.  It can be safely
assumed that the velocity range data for any sample is drawn from the
same population, regardless of redshift, radio size or radio power.   

\begin{table*}
\caption{Spearman Rank Correlation Coefficients. These are given for
the correlations of the maximum
velocity FWHM with redshift, radio power and radio size, along with
partial rank coefficients for the same parameters. Significance levels
for each correlation are also tabulated where these are greater than
85\%. K = FWHM, z =
redshift, P = radio power and D = radio size.  The labelling of the
correlation coefficients is as for Table 1.} 
\begin{center} 
\begin{tabular}{cc@{=}r@{.}lc@{=}r@{.}lc@{=}r@{.}l} 
\hline
Sample&\multicolumn{9}{c}{Kinematics correlations...}\\
&\multicolumn{3}{c}{...with redshift}&\multicolumn{3}{c}{...with radio power}&\multicolumn{3}{c}{...with radio size}\\
\hline
3C at $z \sim 1$ ($n$=14) 		 &$\rm{r}_{\rm{Kz}}$	& 0&079       &$\rm{r}_{\rm{KP}}$   & 0&077       &$\rm{r}_{\rm{KD}}$   &-0&650 (-99.4\%)\\
6C at $z \sim 1$ ($n$=7) 			 &$\rm{r}_{\rm{Kz}}$	&-0&536 (-89\%)    &$\rm{r}_{\rm{KP}}$   &-0&450 (-85\%)    &$\rm{r}_{\rm{KD}}$   &-0&821 (-99\%)\\ 
Radio power matched low redshift 3C ($n$=8)&$\rm{r}_{\rm{Kz}}$	& 0&566 (+93\%)  &$\rm{r}_{\rm{KP}}$   & 0&325    &$\rm{r}_{\rm{KD}}$   &-0&374\\\\
All three subsamples			 &$\rm{r}_{\rm{Kz}}$    & 0&520 (+99.8\%)  &$\rm{r}_{\rm{KP}}$   & 0&570 (+99.94\%) &$\rm{r}_{\rm{KD}}$   &-0&559 (-99.92\%)\\
($n$=29)					 &$\rm{r}_{\rm{Kz|P}}$  & 0&410 (+99\%)  &$\rm{r}_{\rm{KP|D}}$ & 0&541 (+99.88\%) &$\rm{r}_{\rm{KD|z}}$ &-0&464 (-99.4\%)\\
					 &$\rm{r}_{\rm{Kz|D}}$  & 0&541 (+99.88\%) &$\rm{r}_{\rm{KP|z}}$ & 0&227 (+88\%)    &$\rm{r}_{\rm{KD|P}}$ &-0&558 (-99.92\%) \\
					 &$\rm{r}_{\rm{Kz|PD}}$ & 0&299 (+94\%)  &$\rm{r}_{\rm{KP|Dz}}$& 0&298 (+93\%)  &$\rm{r}_{\rm{KD|zP}}$&-0&496 (-99.62\%) \\\\
Including the full low redshift subsample&$\rm{r}_{\rm{Kz}}$    & 0&760 ($>$99.99\%)&$\rm{r}_{\rm{KP}}$   & 0&765 ($>$99.99\%)&$\rm{r}_{\rm{KD}}$   &-0&518 (-99.98\%)\\
($n$=42)				 &$\rm{r}_{\rm{Kz|P}}$  & 0&399 (+99.54\%) &$\rm{r}_{\rm{KP|D}}$ & 0&743 ($>$99.99\%)&$\rm{r}_{\rm{KD|z}}$ &-0&455 (-99.88\%)\\
					 &$\rm{r}_{\rm{Kz|D}}$  & 0&758 ($>$99.99\%)&$\rm{r}_{\rm{KP|z}}$ & 0&168 (+99\%)  &$\rm{r}_{\rm{KD|P}}$ &-0&526 (-99.98\%)\\
					 &$\rm{r}_{\rm{Kz|PD}}$ & 0&337 (+98\%)  &$\rm{r}_{\rm{KP|Dz}}$& 0&255 (+94\%)  &$\rm{r}_{\rm{KD|zP}}$&-0&487 (-99.94\%)\\
\hline
\end{tabular}
\end{center}
\end{table*}

\begin{table*}
\caption{Principal component analysis of emission line gas FWHM and
other radio source parameters. This analysis has 
been carried out on the two $z \sim 1$ subsamples and either the radio
power matched or full low-redshift subsamples.  Other details as Table
2.} 
\begin{center} 
\begin{tabular} {ccr@{.}lr@{.}lr@{.}lr@{.}lcccr@{.}lr@{.}lr@{.}lr@{.}lc} 
\hline
{Variable} & \multicolumn{10}{c}{Eigenvectors for the three matched
samples} && \multicolumn{10}{c}{Eigenvectors including the full
low redshift data}\\ 
&&\multicolumn{2}{c}{1}&\multicolumn{2}{c}{2}&\multicolumn{2}{c}{3}&\multicolumn{2}{c}{4}&&&&\multicolumn{2}{c}{1}&  \multicolumn{2}{c}{2}&\multicolumn{2}{c}{3}&\multicolumn{2}{c}{4}\vspace{2.5pt}\\
FWHM&&0&5554 & 0&1825 & 0&5564 & 0&5904 &&&& 0&5553 & 0&1273 & 0&8200 & 0&0544& \\
z  && 0&5433 &-0&2665 &-0&7470 &-0&2753 &&&& 0&5634 &-0&2382 &-0&2959 &-0&7337& \\
P  && 0&4914 &-0&5362 & 0&3261 & 0&6039 &&&& 0&5442 &-0&3447 &-0&3598 & 0&6750& \\
D  &&-0&3936 &-0&7799 &-0&1613 &-0&4593 &&&&-0&2794 &-0&8990 & 0&3325 &-0&0567& \vspace{2.5pt}\\
{Percentage} && \multicolumn{2}{c}{61.73}& \multicolumn{2}{c}{22.39}&\multicolumn{2}{c}{8.27}&\multicolumn{2}{c}{7.61}&&&& \multicolumn{2}{c}{67.87} &\multicolumn{2}{c} {23.62} & \multicolumn{2}{c}{5.47} & \multicolumn{2}{c}{3.04}&\\
\hline
\end{tabular}
\end{center}
\end{table*}

Another parameter we have studied previously is the luminosity of
the \oo line scaled by the ratio of the radio power of a source to the
mean value for the sample, and its variation with radio power (see
Paper 1). Due to
different observational techniques (long-slit spectroscopy versus
narrow band imaging), a strict comparison between the \oo luminosities
of the low and high redshift galaxies is not feasible.  Nonetheless
there is little variation in scaled $L\rm{_{[OII]}}$ with radio size
for the low redshift sources.  A decrease in scaled  $L\rm{_{[OII]}}$
with radio size as seen for the $z \sim 1$ data is \emph{not}
observed, suggesting that any boosting of line luminosity by shocks in
the smaller sources is much less important at lower redshifts.  This
would be consistent with the less extreme kinematics observed in the
low redshift subsample.  If the kinematic results are due in part to
radio power/shock strength effects, as found from the strong
independent correlation of FWHM with radio power, a weaker or
non-existent decrease in $L\rm{_{[OII]}}$ with radio size would also
be expected.  

\section{Discussion}

In this paper, we have compared the ionization properties of both the 6C and 3CR subsamples at $z
\sim 1$ with the ionization properties of the low--intermediate
redshift galaxies ($z < 0.7$) in the sample of Tadhunter {\it et al}
(1998).  We have shown that, whilst not as effective as the \crat line
ratio, [O\textsc{ii}]/H$\beta$ can be used as a useful diagnostic of
the emission line ionization mechanism.  The [O\textsc{ii}]/H$\beta$
ratios of the Tadhunter sample galaxies at low redshift clearly
display the same variation between the degree of ionization and the
radio source size, or alternatively, source age and ionization
mechanism.  We have contrasted the kinematics of our $z \sim 1$
samples with galaxies with $z < 0.7$ selected from the sample of Baum
\& McCarthy (2000).  The data are strongly correlated with both
redshift and radio luminosity, as well as being strongly
anticorrelated with radio size.  Even considered  individually, the
three subsamples display similar trends, although the kinematics of
the low redshift galaxies are clearly less extreme than their high
redshift counterparts.  In the low redshift objects, the lack of any
anticorrelation between radio size and \oo luminosity, scaled by the
radio power of the host galaxy, suggests a much weaker (or perhaps
non-existent) relation between emission line strength and radio source
age. It also implies that shock induced boosting of emission line
luminosities is much less important in the low redshift sources than
for similar radio power sources (i.e. 6C galaxies) at $z \sim 1$.   

The variations with radio size are apparent across the full range of
redshifts, and have been well documented in previous studies
(e.g. Best {\it et al} (2000b) and Paper 1).  In
general, ionization state, EELR size, velocity profile and FWHM all
vary with radio size.  Kinematic and ionization properties of the
EELRs are found to be intrinsically correlated, despite the selection
biases.  It is clear that photoionization by the AGN is the only
important mechanism for the more evolved, larger radio sources, which
also display relatively undisturbed kinematics.  The influence of
shocks on the kinematics of the EELR gas clouds is certainly of
importance, providing an explanation of the very distorted kinematics
seen in the smaller radio sources.  The clear importance of radio
source shocks in these sources suggests that a gravitational origin
for the observed extreme kinematics is unlikely.   However, in the
case of large, photoionized sources, where radio source shocks are less
effective, the increased FWHM observed at high redshifts
could in fact be due to a larger gravitational potentials,
for example if the radio galaxies are associated with 
clusters (e.g. Best 2000) rather than just the potential of 
an individual galaxy.

We will now consider the dependence of the ionization and kinematic
properties of the EELRs on redshift and radio power independently. 
The inclusion of 6C data in the combined sample weakens the inherent
$P$--$z$ correlation rather than removing it entirely.  However, using
the combined sample to investigate the relationships between the EELR
properties and these parameters does allow the redshift--radio power
degeneracy to be successfully broken.  With this achieved, we can
create a fuller picture of the processes occurring within the emission
line regions of all three subsamples of radio galaxies.  The
ionization state of the gas shows no significant variation with
redshift, and depends only on radio power, in addition to the size of
the radio source.  The interplay between the strength of any shocks
involved and the ionizing UV flux of the AGN is still unclear;
nevertheless, the emission line regions of the powerful $z \sim 1$ 3CR
radio sources generally exist in a higher ionization state than the
sources in the low redshift or 6C subsamples, as shown by the
variation of the observed [O\textsc{ii}]/H$\beta$, \crat and \nrat
line ratios.  The ionization state is clearly dependent on AGN
properties rather than other host galaxy properties or the environment
of the radio source.   

The ionization parameter $U$, which is proportional to the ratio of
ionizing photons impinging on the surface of a cloud to the gas
density, is approximately twice as great for the large 3CR radio
sources as for the large 6C sources, although the distributions of
sources from both subsamples cover a range of values for $U$ ({\it Paper
1}).  Although the difference in ionization parameter is less than the
difference in radio power between the two samples (a factor of $\sim
5$ at $z \sim 1$), this anomaly could perhaps be explained by less
dense clouds populating the EELRs of 6C galaxies compared with their
more powerful 3CR counterparts.   The sensitivity of the observations
could also be an important factor.  For the more powerful 3CR sources
at $z \sim 1$, photoionized gas would be observable out to a greater
distance from the central AGN than for the less powerful 6C sources at
the same redshift.  As the flux of ionizing photons impinging on a
cloud per unit surface area decreases with distance from the AGN, the
ionization parameter for more distant clouds will be lower than that
for clouds closer to the AGN.  The increase in the observed extent of
the EELRs of more powerful radio sources will therefore lead to the
average value of the ionization parameter being lower than that for a
less powerful radio source observed to the same sensitivity.  However,
if the clouds are in pressure equilibrium, their density is also
likely to decrease with distance from the AGN, reducing the variation
in $U$ with distance.   Alternatively, the ionizing photon flux may
scale non-linearly with radio luminosity, as suggested by the work of
Serjeant {\it et al} (1998) and Willott {\it et al} (1998).  They find
that the optical luminosity of quasars scales roughly as
$L_{\rm{rad}}^{0.5}$.   This result suggests that the difference in
ionization parameter for the two $z \sim 1$ subsamples can be expected
to be less than the factor of five difference in their radio powers.

We also observe the effects of radio power on the kinematics of the
emission line regions, which are independently correlated with radio
power once the correlations with other parameters (redshift and radio
size) are removed.  If the disturbed kinematics of smaller radio
sources are due in the main to the influence of shocks, parameters
such as the gas velocity FWHM are expected to be lower for the weaker
shocks associated with less powerful radio galaxies. In addition, the
more powerful radio sources may involve an increased number of
jet--cloud interactions.

Finally, the observed decrease in gas velocity FWHM with redshift
needs to be explained.  For the kinematics, redshift is a much more
important parameter than radio power, as shown by the results of the
partial rank correlations and principal component analysis.  This
result seems to point to variations in either the host galaxy
properties, e.g. mass of gas available in the emission line regions, 
or changes in, say, the IGM density with redshift.  Changes in cloud
sizes and/or densities would clearly affect any acceleration of the
clouds due to shocks associated with the expanding radio source.  
An increase in jet--cloud interactions may be due to a higher gas
density in the ICM for the higher redshift galaxies.  High redshift
3CR galaxies are generally found in clusters, whereas low redshift 3CR
sources are more likely to be located in smaller groups (e.g. Hill \&
Lilly 1991, Best 2000). A change in radio source environment with
cosmic epoch may have an effect on the observed kinematics.    The
more extreme kinematics of the EELRs of higher redshift radio galaxies
can plausibly be attributed in part to their richer environments as
compared with low redshift radio galaxies.  

\section{Conclusions}

Our conclusions are as follows:
\begin{enumerate}
\item A comparison between the emission line diagnostic plots in Paper
1 and the current paper suggests that the shock model with a precursor
model best explains the spectra of small double radio sources.   The
emission line regions of larger radio sources are well explained by
photoionization by the UV flux from the AGN.  An accurate description
of the spectra of small radio sources is likely to require a
combination of both shock ionization and photoionization, even where
shock ionization is likely to be the most important ionization mechanism.   
\item In addition to the known variation in EELR gas kinematics with
radio size, the kinematic properties are strongly correlated with
redshift and radio power independently.   The more extreme gas
kinematics observed at earlier cosmic epochs suggest that the 
structure and/or composition of the EELR gas clouds may vary with
redshift.  Evolution in the environment of the radio source
(e.g. changes in IGM/ICM density profile) may also be occuring.  The
dependence of the EELR kinematics on radio power is likely to be due
to the influence of shocks.
\item The ionization state of the gas is strongly independently
correlated with radio size, as well as less strongly correlated with
radio power.  The data show no dependence on the redshift of the
source once radio power effects have been removed.  The ionization
state, and hence the dominant source of ionizing photons, is therefore only
dependent on the properties of the AGN (i.e. radio size and radio
power) and not on cosmic epoch.  
\item Whilst there is a fairly tight distribution of scaled
$L\rm{_{[OII]}}$ with radio size in the low redshift subsample, the
weak positive correlation of this data suggests that unlike the strong
anticorrelation of emission line luminosity with source age observed
in the redshift $z \sim 1$ sources,  this correlation is much weaker
or non-existent for the lower redshift radio galaxies.  This indicates
that shocks are less important for the radio galaxies at low
redshifts, and is consistent with our other results.
\end{enumerate}

\section*{Acknowledgements}

This work was supported in part by the Formation and Evolution of
Galaxies network set up by the European Commission under contract ERB
FMRX--CT96--086 of its TMR programme.  We would like to thank Mark
Allen for providing digitised \textsc{mappingsii} line ratios.
KJI acknowledges the support of a PPARC research studentship.  PNB is
grateful for the generous support offered by a Royal Society Research
Fellowship. 
The William Herschel Telescope is operated on the island of La Palma
by the Isaac Newton Group in the Spanish Observatorio del Roque de
los Muchachos of the Instituto de Astrof{\'{\i}}sica de Canarias.  
We would like to thank the referee for useful comments.

\label{lastpage}

\end{document}